\documentclass[11pt,a4paper]{article}
\usepackage{naaclhlt2019}
\usepackage{times}
\usepackage{amsmath}
\usepackage{latexsym}
\usepackage{graphicx}
\usepackage{adjustbox}
\usepackage{makecell}

\usepackage{array,booktabs,multirow}
\newcolumntype{C}{>{\fontfamily{fdr}\fontseries{b}\selectfont}c}
\newcolumntype{T}{>{\fontfamily{fdm}\selectfont\small}c}
\newcolumntype{P}{>{\fontfamily{fdm}\selectfont\small}p{3cm}}
\usepackage[T1]{fontenc}
\usepackage{droidserif}
\usepackage{sourcesanspro}

\usepackage{url}

\aclfinalcopy 

\title{An Ontology-driven Treatment Article Retrieval System \\ for Precision Oncology}

\author{Zheng Chen, Sadid A. Hasan, Joey Liu, Vivek Datla, Md Shamsuzzaman, Hafiz Khan,\\ \textbf{Mohammad S Sorower, Gabe Mankovich, Rob van Ommering, Nevenka Dimitrova} \\ \textbf{Philips Research North America, Cambridge, MA, USA}}

\date{}

\begin{document}
\maketitle
\begin{abstract}
  This paper presents an ontology-driven treatment article retrieval system developed and experimented using the data and ground truths provided by the TREC 2017 precision medicine track. The key aspects of our system include: meaningful integration of various disease, gene, and drug name ontologies, training of a novel perceptron model for article relevance labeling, a ranking module that considers additional factors such as journal impact and article publication year, and comprehensive query matching rules. Experimental results demonstrate that our proposed system considerably outperforms the results of the best participating system of the TREC 2017 precision medicine challenge. 
\end{abstract}

\section{Introduction} \label{intro}
Providing useful precision medicine (PM)-related information to clinicians treating cancer patients is in high demand.  For example, given a case with patient's disease (type of cancer), the relevant genetic variants (genes, mutations, alterations), basic demographic information, and other potential factors that may be relevant, the clinician may want to find out biomedical articles addressing relevant treatments for the given patient. The process of manually searching relevant biomedical articles requires lots of significant medical research and investigation based on an underlying clinical scenario, which is often time and effort consuming due to the large volume of online medical articles and the unstructured nature of the documents and texts. To this end, we develop an automated medical article retrieval system to enable clinicians efficiently find the most relevant medical research articles useful for cancer treatments.  

The Text REtrieval Conference (TREC) 2017 offered a PM track for this task, focusing on oncology and genetic mutations of cancer {\cite{roberts2017overview}}. TREC provides a January 2017 snapshot of about 27M PubMed Central (PMC) abstracts and 30 synthetic patient profiles. The patient profiles consist of disease names, relevant gene names and variations, patient demographic information, and other conditions; one example patient profile (a.k.a. topic) is shown below in Table {\ref{table:1}}.

\begin{table}[h!]
\centering
 \begin{tabular}{|l|} 
 \hline
 \textbf{Disease}: Lung adenocarcinoma \\
 \textbf{Genes}: KRAS (G12C) \\
 \textbf{Demographic}: 61-year-old female \\
 \textbf{Other}: Hypertension, Hypercholesterolemia \\
 \hline
\end{tabular}
\caption{TREC topic example}
\label{table:1}
\end{table}

In this paper, we conduct our experiments with the TREC benchmark collection. The most notable aspects of our proposed article retrieval system include: 1) leveraging various disease, gene, and drug ontologies, 2) extensive query formulation rules, 3) a novel perception model for relevance labeling, and 4) a ranking module that considers criteria such as publication recency and journal impact factors. Experimental results demonstrate that our system considerably outperforms the results of the best participating system of the TREC 2017 precision medicine challenge.

\begin{figure*}[h]
  \includegraphics[width=\textwidth]{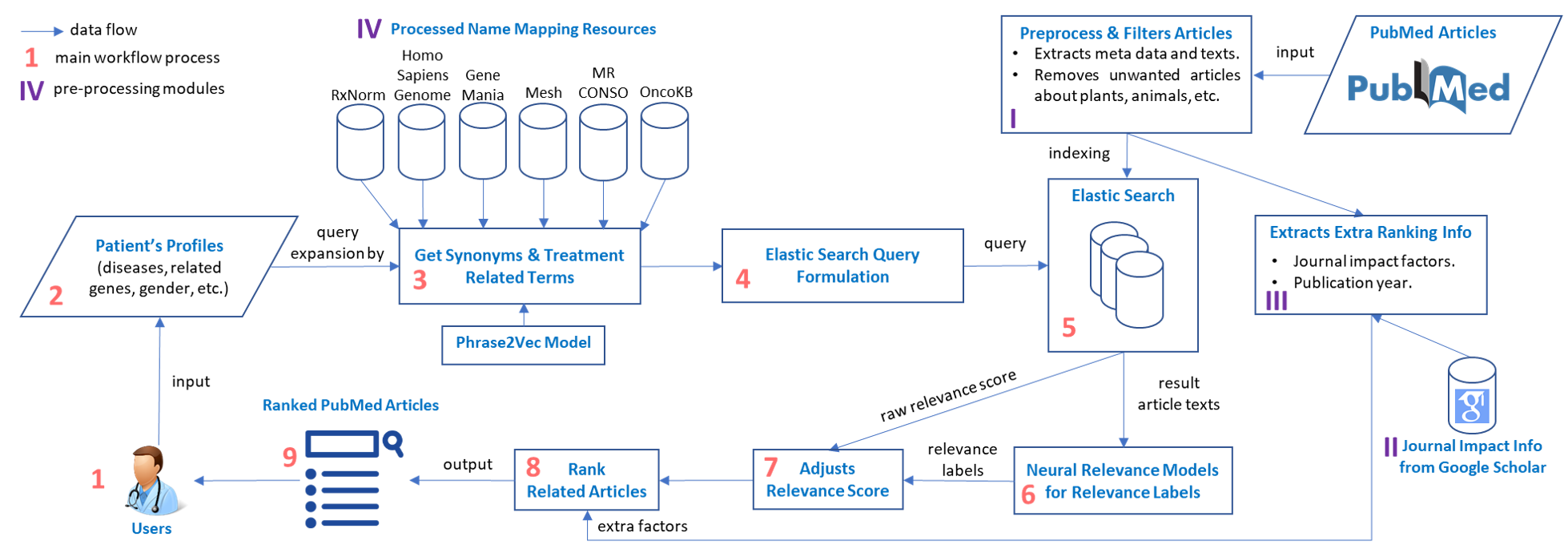}
  \caption{The pipeline of our treatment-related biomedical article retrieval system.}
  \label{figure:1}
\end{figure*}

\section{System Description} \label{desciption}
We build an end-to-end pipeline for retrieving relevant treatment-related biomedical articles given a patient profile. The pipeline has four pre-processing modules and nine main-cycle processes, as shown in Figure {\ref{figure:1}}.

\subsection{Preprocess} \label{preprocess}

The pre-processing modules are responsible for clean the PubMed abstract data, the ontology data and ranking resources, and then index these data in appropriate databases to make them readily for use in the main retrieval process cycle.

\textbf{\textit{PubMed abstracts}}. The TREC 2017 PM track clearly stated the retrieval focuses on oncology {\cite{roberts2017overview}}, and the thirty patient profiles are all for human. Therefore, we would like the pre-process to filter out non-cancer or non-human treatment articles. To achieve this, we notice each article in the PubMed abstract data set comes with a MeSH code. The MeSH code system {\cite{meshlist}} is a comprehensive controlled vocabulary for the purpose of indexing journal articles and books in the life sciences. We thus only keep articles whose MeSH codes starting with either "C" or "D", corresponding to the human "diseases" and the "chemicals and drugs" category, and index them in an \textit{Elasticsearch} database. The vast majority the cancer-related and human-treatment-related articles are under the two categories.

\textbf{\textit{Ontologies}}. One distinct feature of our retrieval system is the extensive inclusion of different ontologies. The TREC patient profile itself is insufficient to match with the database. The diseases and genes might be represented by different terms in different publications. Moreover, we believe the existence of a related drug name is a significant indicator of a relevant treatment article. These information is not present in the TREC patient profile, and our  employment of the following ontologies to expand user queries is crucial to the retrieval performance.

\begin{figure*}[h]
  \includegraphics[width=\textwidth]{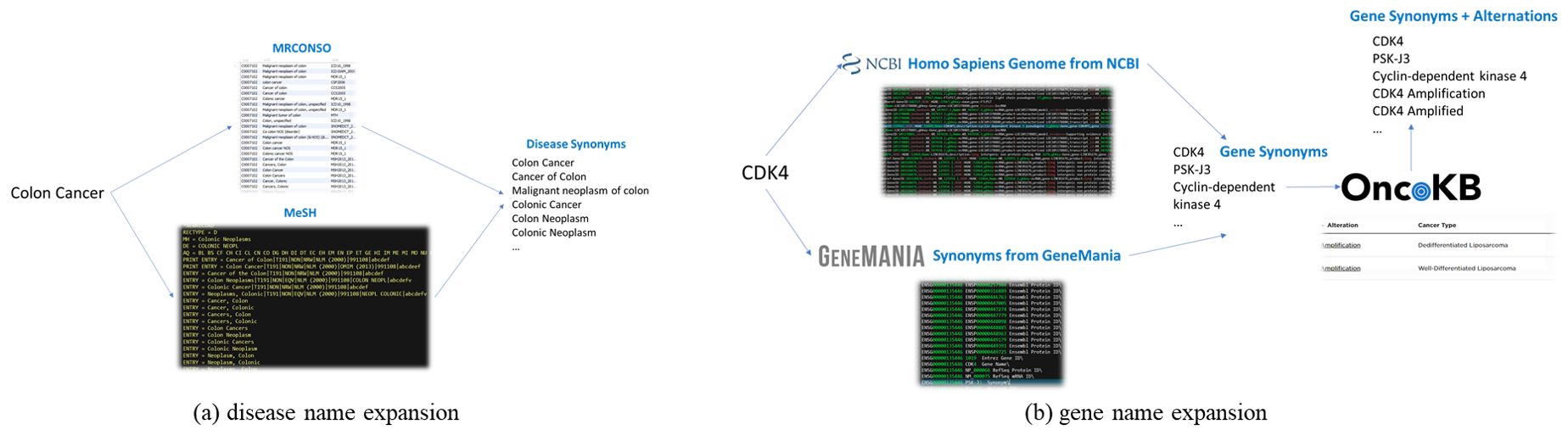}
  \caption{(a) a disease name matches with ontology MRCONSO and MeSH to find its synonyms/aliases; (b) a gene name first matches with gene ontologies to find their synonyms/aliases, which further match with OncoKB to find the gene variants.}
  \label{figure:2}
\end{figure*}

\begin{itemize}
    \item For diseases, we built a unified ontology named \textit{MRCONSO} (TODO: the meaning of this name), merging UMLS {\cite{bodenreider2004unified}}, ICD0 and ICD10 {\cite{percy1990international}}, etc. In MRCONSO, different terms for the same disease are mapped to a unique identifier so we know what terms are synonyms. In addition, the 2018 MeSH data file {\cite{meshlist}} is also imported to our database; some disease name aliases are available in this ontology.
    \item For genes, we imported the gene data from NCBI and GeneMannia. {\cite{warde2010genemania}}. Both ontologies contain synonym and acronym information for human genes. For example, with the help of both ontologies, we are able to identify "Cyclin Dependent Kinase 4", "Cell Division Protein Kinase 4", "CDK4", "PSK-J3", "CMM3" are all for the same gene.
    \item For gene variants, we uses the OncoKB {\cite{chakravarty2017oncokb}}. The latest version has more than 4,000 unique mutations, fusions, and alterations for nearly 600 cancer associated genes. We'll later show OncoKB makes noticeable performance improvement for the retrieval. 
    \item For drug names, we applied the phrase2vec model in {\cite{hasan2015using}} trained with the PubMed data to infer drug names from RxNorm {\cite{liu2005rxnorm}} that are related to the provided diseases and genes. In addition, OncoKB also lists related drugs for some variations.
\end{itemize}

\textbf{\textit{Ranking resources}}. Currently the only ranking resource used by our system is the journal impact factors. We collected the H5 indexes of life science related journals from Google Scholar at \url{https://scholar.google.com/citations?view_op=top_venues&hl=en}. This data enable our system to promote articles published in better journals.

\subsection{Retrieval}

Our system now allows a clinician to input a patient profile by entering disease names, gene names, demographic information (e.g. gender, age, race), and other conditions, etc. into the user interface. For the evaluations in this paper, our system directly reads the TREC patient profiles from files. After that, our system goes through several steps, starting with \textit{profile expansion}, and then \textit{query formulation and matching}, and then it does the \textit{relevance labeling} \textit{relevance score adjustment} and \textit{ranking}, and finally it displays the results to the user interface.

\textbf{Profile expansion}. The system searches for synonyms or aliases of the disease names and gene names in the indexed ontologies described in section \ref{preprocess}. All available gene variants are retrieved from OncoKB for the input genes. If a gene variant is not specified in the patient profile, then these variants will later be equally considered in the Elasticsearch query; otherwise, we decide the retrieval should primarily consider the specified variants. The system also identifies drug names related to the input diseases and genes from RxNorm. An illustration of profile expansion is shown in \ref{figure:2}.

\textbf{Query formulation and matching}. Our system generally formulates elastic search queries according to heuristic rules following the TREC 2017 relevance guidelines mentioned in section \ref{intro}.
\begin{itemize}
    \item The retrieved article must contain ALL of the following,
        \begin{itemize}
            \item one disease name or one of its synonyms/aliases.
            \item one gene name or one of its synonyms/aliases.
            \item one gene variant if it is specified.
            \item one drug name related to the disease and gene, or one of some predefined treatment keywords like "treatment", "surgery", "therapy", "radiotherapy", "immunotherapy", "chemotherapy", etc.
        \end{itemize}
    \item The returned article must not contradict with the demographic requirement. For example, if the patient profile specifies "male", then a retrieved article must either be about male, or not specifying a gender
    \item The retrieved article optionally matches the following; if matched, a higher Elasticsearch relevance score will be assigned to the retrieved article.
        \begin{itemize}
            \item gene variants, if none is specified.
            \item other conditions.
        \end{itemize}
\end{itemize}

An Elasticsearch query is formulated by above rules, and is then sent to the Elasticsearch indexing system to match relevant articles. Each retrieved article is assigned a relevance score by Elasticsearch, which is calculated a hybrid of several relevance measurements tf-idf, Coord and Length Norm {\cite{gormley2015elasticsearch}}.

\subsection{Relevance Labeling}

Our system currently employs a Perceptron with frequency features as its input. The frequency features include bag-of-word and phrase frequencies of disease names, gene names, drug/treatment names in title, abstract and keywords respectively. We experimented using three optimizers (SGD, Adadelta, Adagrad) and 5 levels of learning rates ($1$ to $10^{-5}$) are tested, and found Adadelta with learning rate $0.01$ has the best performance. We use this perception for binary classification if each label is relevant or irrelevant. Of course, some irrelevant articles might be incorrectly labelled; we then depend on the relevance score and the ranking to prevent irrelevant articles from entering the top results. Our experiment in \ref{evaluation} shows this Perceptron is beneficial to the retrieval performance.

\subsection{Result Ranking}

A real-life retrieval system should consider factors other than relevance, even though it is not mentioned in TREC relevance guideline. For treatment article retrieval, it is essential to consider publication recency and journal impact factors, as the oncology clinicians would highly expect recent treatment methods published in good journals. The principal of our article ranking is that the adjusted relevance score is still the primary ranking factor, but for articles with close relevance scores, other factors come into play. By this idea, we define a "closeness" parameter $k$, and let $s$ be the adjusted relevance score as discussed above, then the primary ranking score of our system is $r_1=\lfloor s/k \rfloor$ $(\lfloor \cdot \rfloor$ denotes a floor function). After this operation, some articles with close enough adjusted relevance scores (within distance $k$) will have the same primary ranking score. Currently our system uses $k=20$.

These extra factors are often of different scales. For example, the journal impact factor can range from 0 to hundreds, and the publication year for our data set can range from 1950 to 2018. Therefore, we first normalize all extra factors to the range $(0,1)$ by a shifted function $\sigma(x,\tilde{x})=\frac{1}{1+e^{-c(x-\tilde{x})}}$ where $x$ denotes the factor to normalize, $x=\tilde{x}$ is its symmetry axis, $c$ is the horizontal scale of the sigmoid curve, and $\tilde{x}$ can be interpreted as a threshold, and $c$ decides how fast the score increases when the factor $x > \tilde{x}$, and symmetrically how fast the score decreases when the factor $x < \tilde{x}$. If the value of the factor $x > \tilde{x}$, then it results in a score higher than 0.5. For example, let $y$ denote the publish year of a journal, and let $\tilde{y}=2008$, $c=1$, and $\sigma(y,\tilde{y})=\frac{1}{1+e^{-(y-2008)}}$) will give a score lower than 0.5 for articles before 2008, and a score higher than 0.5 for articles after 2008. After normalization, we calculate the secondary ranking score as 
$$r_2 = \frac{w_s(s \text{ mod } k)}{{k} + w_h\sigma(h,\tilde{h})} + w_y\sigma(y,\tilde{y})$$
where $h$ denotes the journal impact factor, $y$ denotes the article publication year; $\tilde{h},\tilde{y}$ are the symmetric axes as defined above, and we use $\tilde{h}=200$, $\tilde{y}=2008$; $w_s,w_h,w_y$ are weights for each corresponding factor.

\begin{table*}[ht]
\begin{adjustbox}{width={\textwidth},totalheight={\textheight},keepaspectratio}%
\begin{tabular}{cccccccc} \toprule
    {} & \thead{OncoKB \\ + Ranking \\ + Perceptron} & \thead{OncoKB \\ + Perceptron} & \thead{Ranking \\ + Perceptron} & \thead{Ranking \\ Only} & \thead{Our \\ Previous$^1$} & \thead{TREC Best} \\ \midrule
    \thead{P5}  & \begin{tabular}{@{}c@{}} 0.7800 \\ (+37.6\%)$^2$ \\(+11.4\%)$^3$ \end{tabular} & \begin{tabular}{@{}c@{}} 0.7767 \\ (+37.1\%) \\ (+11.0\%) \end{tabular} & \begin{tabular}{@{}c@{}} 0.7533 \\ (+33.0\%) \\ (+5.0\%) \end{tabular} & \begin{tabular}{@{}c@{}} 0.6867 \\ (+21.2\%) \\ (-1.9\%)  \end{tabular} & \begin{tabular}{@{}c@{}} 0.5667 \end{tabular} & 0.7000$^4$ \\ \midrule
    \thead{P10}  & \begin{tabular}{@{}c@{}} 0.7333 \\ (+38.4\%) \\ (+14.6\%) \end{tabular}  & \begin{tabular}{@{}c@{}} 0.7333 \\ (+38.4\%) \\(+14.6\%) \end{tabular} & \begin{tabular}{@{}c@{}} 0.6967 \\ (+31.5\%) \\ (+8.9\%) \end{tabular}  & \begin{tabular}{@{}c@{}} 0.5567 \\ (+5.0\%) \\ (-13.0\%) \end{tabular} & 0.5300 & 0.6400  \\ \midrule
    \thead{R-prec}  & \begin{tabular}{@{}c@{}} 0.3671 \\ (+41.5\%) \\ (+22.7\%) \end{tabular}  & \begin{tabular}{@{}c@{}} 0.3658 \\ (+39.5\%) \\ (+22.2\%) \end{tabular} & \begin{tabular}{@{}c@{}} 0.2866 \\ (+9.3\%) \\ (-4.2\%) \end{tabular}  & \begin{tabular}{@{}c@{}} 0.2798 \\ (+6.7\%) \\ (-6.5\%) \end{tabular}  & 0.2622  & 0.2993 \\ \midrule
    \begin{tabular}{@{}c@{}} \thead{NDCG \\ (Top10)} \end{tabular}  & \begin{tabular}{@{}c@{}} 0.2261 \\ (+43.8\%) \end{tabular} & \begin{tabular}{@{}c@{}} 0.2252 \\ (+43.3\%) \end{tabular} & \begin{tabular}{@{}c@{}} 0.2098 \\ (+33.5\%) \end{tabular}  & \begin{tabular}{@{}c@{}} 0.1776 \\ (+13.0\%) \end{tabular} & 0.1572 & \\ \midrule
    \thead{NDCG}  & \begin{tabular}{@{}c@{}} 0.4811 \\ (+21.8\%) \\ (-5.6\%) \end{tabular}  & \begin{tabular}{@{}c@{}} 0.4805 \\ (+21.0\%) \\ (-5.8\%) \end{tabular} & \begin{tabular}{@{}c@{}} 0.4354 \\ (+9.7\%) \\ (-14.7\%) \end{tabular}  & \begin{tabular}{@{}c@{}} 0.4354 \\ (+9.7\%) \\ (-14.7\%) \end{tabular} & 0.3970  & 0.5103$^4$  \\ \midrule
    \thead{infNDCG} & \begin{tabular}{@{}c@{}} 0.4724 \\ (+16.1\%) \\ (+1.7\%) \end{tabular} & \begin{tabular}{@{}c@{}} 0.4719 \\ (+16.0\%) \\ (+1.5\%) \end{tabular} & \begin{tabular}{@{}c@{}} 0.4521 \\ (+11.1\%) \\ (-2.7\%) \end{tabular} & \begin{tabular}{@{}c@{}} 0.4521 \\ (+11.1\%) \\ (-2.7\%) \end{tabular} & 0.4070 & 0.4647 \\ \bottomrule
\end{tabular}
\end{adjustbox}
\caption{Evaluations of our latest retrieval system. }
{\small 1. The result for pms-run5-abs shown in \cite{roberts2017overview}; 2. this percentage is in comparison with our previous result pms-run5-abs; 3. this percentage is in comparison with the best TREC 2017 result; 4. we obtained the results of UTDHLTFF, which performs best in TREC 2017, and evaluate these metrics.}
\label{table:2}
\end{table*}

\section{Experiments}

\subsection{Dataset}

The original PubMed data set provided by TREC has more than 27M abstracts. After the pre-process as described in \ref{preprocess}, a subset of 394,862 articles are indexed in the Elasticsearch. TREC made a ground truth consisting of relevance labels for 21,221 PubMed articles that are manually curated by oncologists. A label rates a PubMed article as relevant by integer $2$, or partially relevant by integer $1$ or irrelevant by integer $0$. The relevance is judged by the TREC 2017 relevance guideline. Our subset of PubMed articles indexed in Elasticsearch includes all ground-truth articles.

\subsection{Evaluation Results} \label{evaluation}

The performance evaluation is conducted using the standard trec\_eval tool at \url{https://trec.nist.gov/trec_eval/}. For a retrieval system, the top results are most important; therefore, we compare metrics like precision at 5/10 and NCDG at 10. We also compare metrics including R-precision and NDCG. The results are shown in Table \ref{table:2}. The first column is the results when all featurs of our system are on; in column 2,3,4, some features are turned off to see how much impact they have on the performance: 1) column 1 and column 2 show the ranking slightly improves the performance; 2) column 1 and column 3 show the OncoKB ontology brings noticeable enhancement; 3) column 3 and column 4 show the effectivenss of the perceptron relevance labeling.

\bibliography{naaclhlt2019}
\bibliographystyle{acl_natbib}

\end{document}